\documentclass[10pt, conference, final]{IEEEtran}
\pdfoutput=1
\usepackage{pslatex}
\usepackage{epsfig}
\usepackage{psfrag}

\begin{document}

\title{Distributed Discovery Clients for Spectrum Allocation }
\author{ Torleiv Maseng$^{1}$, {\O}ivind Kure$^{2}$ and Magnus Skjegstad$^{3}$ 
\\1. Norwegian Defence Research Establishment (FFI)
\\ Postbox 25, 2027 Kjeller, Norway
\\2. ITEM, NTNU
\\3. Cambridge University}

\maketitle

\begin{abstract}
By using a distributed P2P system where the agents reside in in microprocessors already present in most radio nodes like Wi-Fi access points, base stations,  TVs connected to Internet etc, these agents can discover other agents over the back-haul network. As a result, client 
node lists (similar to neighbor lists used in 3GPP) are created.  An alternative is to use a centralized database. Why this best is done by distributed agents is discussed in this paper along with security considerations. Examples of other applications which will benefit from this system  is also presented.

\end{abstract}

\section{Introduction}

In 2010 it was 5 billion connected devices and  in  2020 there will be 50 billions of devices  connected  to Internet \cite{METIS} by radio. This enormous increase is partly caused by introduction of IoT and M2M devices which communicate over short distances, e.g. inside a house. Since databases do not scale with number of nodes and since there is little willingness for many cheap devices to pay for centralized database assistance, we claim that the use of a distributed architecture is better suited. A database may however be used as a supplement.

Once the Distributed Agents are universially implemented, they will simplify coordination among the various device types like Wi-Fi,  power meters, TVs etc communicating via IP packets over the back haul network. These  agents can be the same for all these devices, but the    Dynamic Spectrum Allocation (DSA) will most likely depend on the radio system and  tailored specifically to each system. This DSA application will enable the different radio devices to share the same frequency band  while maximizing capacity of the radio links by minimizing interference between nodes operating in the same band.

Cognitive Radio (CR) or DSA requires control channels to set up the radio channel. Since these  are  not standardized for all type of radios and require different radio protocols, we propose to use the back haul network or the Internet for control channels since this is available to most access points, hubs, base stations etc. We have previously proposed a decentralized DSA architecture and P2P protocol \cite{skjegstad14}. In this paper, we argue that decentralized DSA improves the radio capacity for IoT and M2M communication in the open spectrum. We also discuss security implications and future directions of research. 

DSA is different for IoT and M2M devices compared to the process for a mobile cellular operator. The local  IoT and M2M nets are mainly owned  by individuals and produced by companies whose prime business is to produce components and not run network management. Therefore the structure of these nets is distributed as opposed to the mobile networks which is centrally controlled. In a mobile network all base stations and their location are known  and the coordination between the base stations is therefore easier. In addition, they are all connected through the back haul network. The coordination proposed in this paper is therefore best suited for IoT and M2M nets unless the mobile operators need to carry out coordination between different mobile operators or between mobile networks and IoT and M2M nets. 

In the case of CR for White space, the IEEE 1900 DYSPAN group defines how the procedures for secondary users to get access to the TV part of the spectrum. ECMA 392 is designed for shorter distances and personal devices operating in the White Space.
IEEE 802.11af was formed for the purpose of wireless local area network (WLAN) operating in TV white space spectrum in the VHF and UHF bands between 54 and 790 MHz which increases the possible range and data rate compared to 802.11 a/b/g/n/ac.  The upcoming IEEE 802.11ah standard which will use the 900 MHz band is expected to be approved by January 2016. That makes 2014-2015 the time frame for development of new silicon for end nodes and access points. The standard is expected to cover many modest home uses at 10-20 Mbits/s.

The IEEE 802.19 is the Wireless Coexistence Technical Advisory Group deals with coexistence between unlicensed wireless networks. Reusing existing spectra assignments has been announced by the Radio Spectrum Policy Group assisting the European on radio spectrum policy issues in EU. Therefore the Norwegian Post and Telecommunications Authority and the Swedish Post and Telecoms Authority argue that DSA and CR should be allowed secondary users as long as they maintain knowledge about other users in their vicinity and do not disturb the primary users. This decision is expected to apply for all future spectrum assignments \cite{nptstrategy}.

Operation of a centralized database requires an organization that is willing to fund and coordinate the location database for all involved receivers and transmitters.  For  Wi-Fi access and IoT and M2M devices such an authority is not immediately identifiable. An alternative is a more organic method based on voluntary participation in peer to peer(P2P) networks. In  \cite{skjegstad14} a Discovery system architecture  is described which enable radio nodes to be discovered using a distributed P2P agent network communicating over the back-haul network. This system is based upon that all the nodes know their own position. 

How positions are obtained for IoT and M2M networks, is discussed in \cite{Positioning} \cite{location} and it can be done in the same way as for UMTS (3G) and LTE . For networks with very short range, note that in order to carry out DSA among them, it it is the radio connectivity matrix with path-losses that is needed and their relative locations are less important. The connectivity matrix may be found by coordinated transmit and receive sessions and from the resulting measurements.

\begin{figure*}
\centering
\includegraphics[width=1\columnwidth]{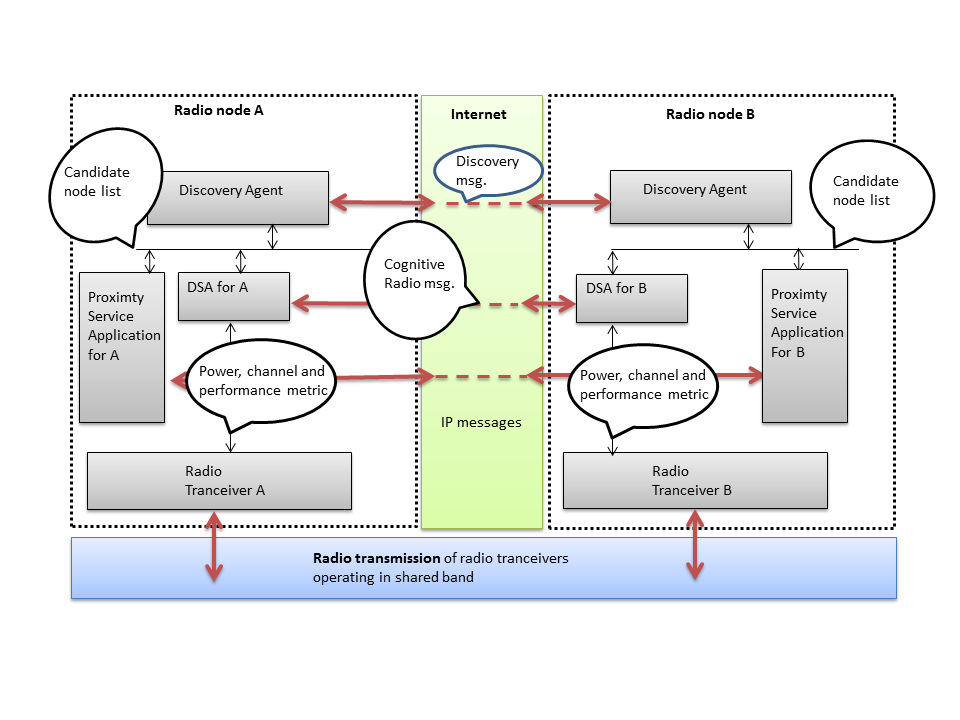}
\caption{Two radio nodes A and B are are connected via the Internet. They both share the same frequency band even if they may use different radio protocols. Once the Discovery Agent is in place, this enables the establishment of Neighbor Relation Tables in LTE and called Client Node Lists in this paper. This list is needed to execute DSA and proximity services application programs which may be different for A and B since they use different radio protocols. In order to make the the Discovery Agents interoperable, the format of the "IP Discovery messages" and their content, need to be defined.}
\label{standardisering}
\end{figure*}

Assuming that the Discovery Agents (DA) reside inside the existing microprocessors in all participating radio nodes, additional features may be offered to the users of radio nodes like Proximity Services in Figure \ref{standardisering}. Examples of these services are 
context-aware services for roaming nodes and for nodes belonging to different organizations, handover and a network consisting of many small base stations.
 
\section{A distributed P2P discovery system}
\label{sec:disco}

In \cite{skjegstad14}, the P2P protocol for discovering radio devices was first published in detail. The objective of the protocol is  to discover radio devices that are operating on the same frequencies in overlapping geographical areas. The P2P network is established over the back-haul network to enable nodes to discover each other without having radio contact.  In order to make the the Discovery Agents interoperable, the format of the "IP Discovery messages" and their content, need to be defined.

The P2P protocol creates an unstructured overlay that connects nodes that are likely to be interfered or interfere with each other. In addition, nodes maintain connections to a few nodes that are far away, so that they can help reduce the discovery time for new nodes that enter the network. The goal of the protocol is to produce a list of network addresses to other nodes one needs to coordinate frequency use with. This is accomplished using two main mechanisms that run on each node.

The first mechanism is a random peer sampling mechanism \cite{jelasity07} that aims to produce a random set of nodes from the overall topology. The random set of nodes is mainly used as a starting point for the discovery mechanism. The random set is also useful to new nodes that have yet to find other nodes from their area yet.

The second mechanism maintains a set of nodes it has discovered so far that have the highest utility according to a utility function, using an adapted version of the T-Man protocol \cite{jelasity09}. The utility function in \cite{skjegstad14} is based on degree of overlap between the frequency coordination areas of the nodes, as we describe later in this section. The more likely two nodes are to interfere each other, the more important they should be considered to be to each other. The mechanism contacts nodes with high utility to exchange information in the hope that they may know about even more important nodes that it is interested in.

Initially, when a node joins the P2P network, the nodes it knows about so far are in the random set of nodes produced by the first mechanism. However, as the second mechanism begins to improve its results by contacting nodes with high utility, we gradually learn more about nodes that are more important. Eventually, we discover the nodes that we need to coordinate frequency use with. These nodes will also have gathered information that we are interested in if they reside in the same area, reducing the overall discovery time.

The protocol has been extensively evaluated though simulation, and its ability to converge and find all nodes even if the starting point (seed) was far away from the location of interest. This was demonstrated in a network with 2.3 million nodes representing Wi-Fi routers corresponding to every household in Norway. The protocol was implemented  in real Wi-Fi routers running the open source Wi-Fi router firmware OpenWRT and made available for free use as long as the origin of the sources code producer is retained in the source code \cite{source}.

Table \ref{tab:newsitem}  lists the data fields distributed with the P2P protocol in \cite{skjegstad14}. The fields should contain enough information to enable the nodes to contact each other and maintain the P2P overlay. As the amount of information included in the fields distributed by the discovery protocol affects the overall bandwidth requirements, it should be kept as small as possible. The total size of the fields in Table \ref{tab:newsitem} is 56 bytes, but could in many cases be reduced further. After the nodes have found each other additional information can be exchanged (such as radio parameters, sensing information etc) directly using the IP address provided by the discovery protocol.

In order to make the Dsicovery Agents interoperable, the format of the “IP Discovery Messages” and their content should be defined.

\begin{table}
	\centering
	\begin{tabular}{|p{2cm}|p{1cm}|p{3.5cm}|}
	\hline
        \emph{Field} & \emph{Length} & \emph{Description} \\ \hline
		Identifier & 8 bytes & Overlay node ID \\ \hline
		Location & 16 bytes & Geographical location \\ \hline
		Coordination radius & 8 bytes & Radius of coordination area \\ \hline
		IPv4 or & 4 bytes & Source IP \\ 
		IPv6 address & 16 bytes & \\ \hline
		Timestamp & 8 bytes & When the news item was created \\ \hline
	\end{tabular}
	\caption{Fields included in "IP Discovery messages". Total length with IPv6 is 56 bytes \cite{skjegstad14}.}
	\label{tab:newsitem}	
\end{table}

Each P2P agent uses the P2P network to gather the network addresses of nodes operating in its surroundings. This is accomplished by periodically contacting other P2P agents over the back-haul network and requesting information about nodes they have discovered. It is based upon a Gossip protocol \cite{jelasity09} and communicates with only one other node at the time as opposed to multicast. Therefore the overhead bandwidth usage is limited. With a periodic interval
of 15 seconds, the average amount of data sent and received from a single client would be approximately 0.5 kilobytes per second.

\begin{figure}
\centering
\includegraphics[width=0.7\columnwidth]{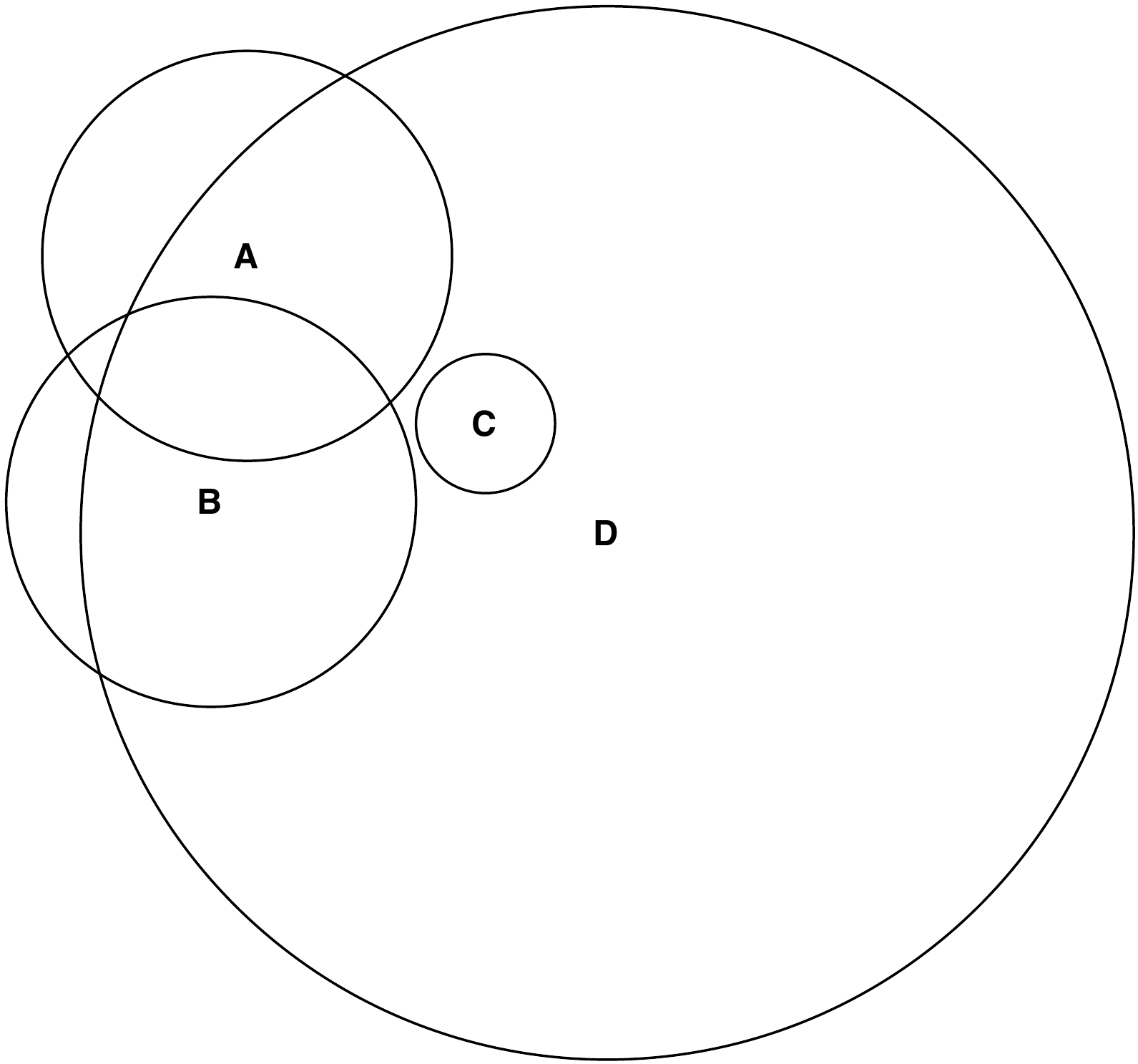}
\caption{Nodes A, B, C and D with coordination areas \cite{skjegstad14}.}
\label{fig:f_example}
\end{figure}

An illustration of how coordination areas are used by the P2P protocol with omnidirectional antennas is shown in Figure \ref{fig:f_example} (from \cite{skjegstad14}). Each node defines a coordination area around itself based on its transmission power and where it has clients. Each node is then interested in discovering other nodes that have coordination areas that overlap their own. The most important nodes (according to the utility function) will be nodes that have the largest overlap between coordination areas. In the illustration, node D will for example be of interest to all nodes as it has a large coordination area that overlaps with them. C on the other hand, may only be of interest to D as it can interfere with its users. Similarly, C may be very interested in getting in contact with D to negotiate how to avoid interference. A and B are mainly interested in each other and D. 

The results are ranked according to relevance and the most relevant nodes are then contacted in the next periodic exchange. Gradually all nodes will discover the nodes that are most relevant to them. We refer to the most relevant nodes as \emph{candidate nodes}, as they are candidates for additional resource negotiation.

In the simulations, the number of exchanges required for 2.3. million nodes to find all their candidate nodes after a new node has joined was  20 on average, corresponding to 5 minutes.

\section{Improving spectrum capacity}

By choosing operating frequency and transmit power wisely, the capacity of radio systems can be improved by minimizing the disturbance between the radio nodes.  This is normally done by using a database in which knowledge of where the TV transmitters are located (PAWS). This spectrum can be further increased if knowledge where some of the TVs are located \cite{6214368} \cite{PAWS}.
In order to reduce interference, it is useful to change the operating frequency of wireless access points. This feature is already present in IEEE Std 802.11™-2012, but for another purpose.

A commonly used criterion for Dynamic Spectrum Allocation systems is to maximize joint capacity. This is both a tough theoretical challenge (NP hard) and an impractical optimization criterion since the allocation may result in some users getting too little and some getting too much capacity. Besides, the parameters needed like path loss and transmit power are hard to measure accurately . There are few parameters which can be measured accurately in most communications systems considering that traffic load, propagation conditions and user location is changing over time. There are many smart algorithms for assigning power and frequency between radio terminals which share the same band  \cite{1542680} \cite{DBLP:journals/corr/abs-1212-0724} .

\section{Practical considerations}
\label{Practical}
The framework has been described with a one to one mapping between IP address and access points location. In an actual deployment there will have to be agents that represent multiple access points. Most enterprises protect their access points behind firewalls. To avoid opening the firewall and use public addresses for each access point, it is better to have one agent that represents all the access points behind the firewall. To enable incoming messages from other agents to enter the firewall, it will be necessary to open a port, using port forwarding.   The operation of the  common protocol will be executed on one machine rather than on many. Several discovery messages will originate from the same agent.

Similarly, IP address will not be used as the actual identifier. Since a large fraction of access points are in private address spaces, their actual IP addresses are not globally unique. A possible solution is either to use the agent IP address concatenated with a local unique identifier, or the mac address of the access point or using port forwarding. In these cases, the information fields in the protocol must be extended with the IP address of the agents, so a mapping between mac address or port number and agent exists.
In order to be self-organizing in private address spaces, an additional discovery protocol between the agent and the access point must be deployed. However, there are no special requirements, and a simple broadcast response protocol is sufficient.

\section{Security Issues and Privacy Issues}
\label{security}

In a system using a licensed spectrum, e.g. UMTS, the spectrum owner has the right to enforce an authentication procedure and has the means to do so, using the SIM card. The spectrum owner has the right to deny to user from participation unlike a user of an unlicensed band like the ISM bands. For a user of an unlicensed band, only the equipment need to adhere to some specification and the user is free to operate. To enforce authorization is therefore difficult. To encourage coopertion only with devices authorized by a third party like Google, would however be a possibility in an attempt to isolate nonauthorized users.

A distributed system has a slight edge against a centralized system regarding the ability to detect misbehaving participants because the participants can compare the information against their own measurement locally and thereby judge the validity of the information. Such a system can be augmented with a distributed reputation system, adding some improvements in the validity checking of the information. In a centralized architecture it is more complex to marshal and collect the relevant measurements. This is because anomalies must me checked for all participants remotely and more information must be transferred to a central point instead of being processed locally.

The vulnerability and possible defense methods of a centralized system against Distributed Denial of Service attacks are well documented in the last decade.  A distributed system is more resilient. A substantial attack can be viewed as an increase in transaction volume. In our simulations, the system scaled well up to millions of nodes. An attack through increasing the number of nodes should therefore have minimal effect. It may stop individual nodes, and make allocation worse that optimal, but not stop the overall system from working.

A possible attack vector is adding and withdrawing access nodes, called churn in the following. An analysis of churn \cite{skjegstad14} showed a reasonable stability for our proposed system. However, large volumes of adding and deletion of nodes will affect the convergence and result in that some nodes will be false and som not yet detected. The allocation may therefore not be correct. Essentially, the lack of convergence will result in hints that potentially will not be worth following and therefore neglected by the access points. As such excessive churn can be used to neutralize the system, but then the scale will be sufficient to be detected by the traditional mechanisms that are already implemented in the back-haul network.  However, individual agents and access points may be attacked and overwhelmed by churn and volume of messages.  This is a risk all nodes in the internet are already running. 

\begin{figure}
\centering
\includegraphics[width=0.95\columnwidth]{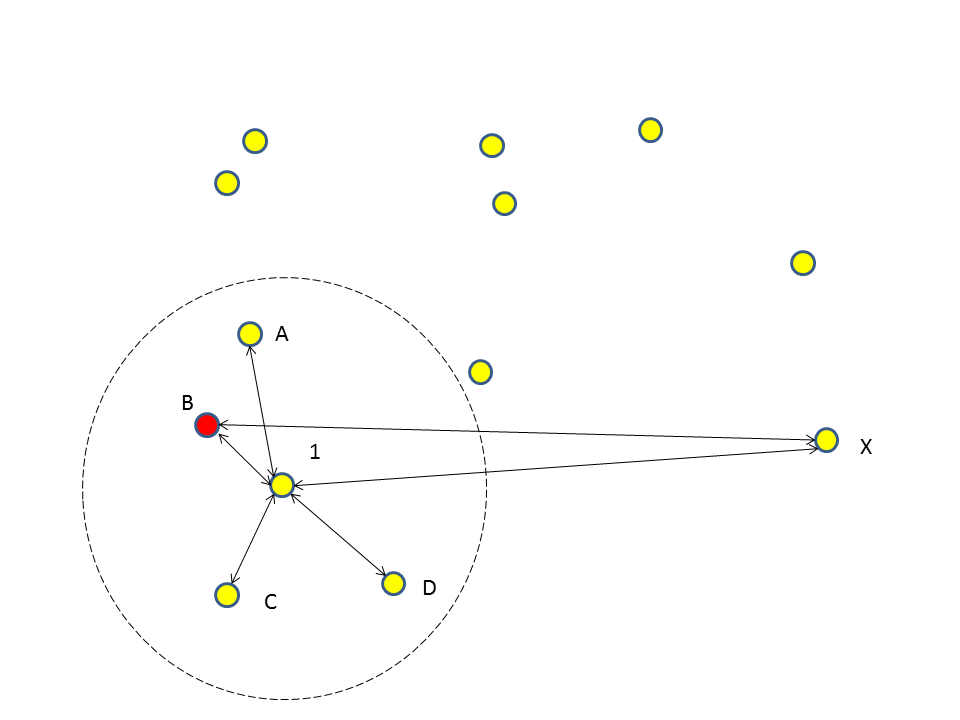}
\caption{Node 1 in the figure has obtained a Neighbor List consisting of node A,B, C and D. In his Neighbor List the IP address of node X is visible at location B. It is node X who runs the agent on behalf of B. This secret is only known to X and B.}
\label{privacy}
\end{figure}

\subsection{Privacy Issues}
\label{privacyissues}

The proposed system raises some privacy issues. In principle, it maps, an IP address range to a particular location.   This information is available to anyone pretending to be at the location and  participating in the system. To hide this information, the system need to create a mapping between access points  and agents that can do the negotiations. This can be done if all users who "opt for privacy", called  "privacy nodes" in the following, must offer to run  agents on behalf of other nodes for a certain period of time. Each privacy nodes select a partner from this pool of nodes randomly. This is illustrated in  in Figure \ref{privacy}. This selection can work asynchronously among the privacy nodes and is selected among privacy nodes with spare capacity to run the agent remotely.
An agent can be selected to represent several nodes and must announce it's current capacity to run a agent on behalf of others nodes at random locations.  The discovery, convergence and negotiations will be the same. However, the IP address of the access point cannot be inferred by the IP address of the agent. The detailed design of such a system  has not been implemented.


%

\subsection{Susceptibility to malignant manipulation  without any central authority}
\label{nocentral}

Some users may invent additional access points  with location close to their actual location using the same channel as the actual point. To external viewers, the channels will appear as overloaded, while in reality it is used by only one access point. Such an attack will not differ from the legitimate use of an agent announcing multiple access points. 
Our recommended solution is to treat the proposed allocation as a hint.  If the system is not manipulated, the outcome is  a “better” solution. However, the hints may be the result of manipulations. In additions, there might also be access points that do not participate in the protocol or refuse to follow the hints. The user must therefore implement their own control strategy for the validity of the hint.    Such a strategy can be implemented without the information exchange with neighbors. 

As an example, the users of the access point can select to follow the hint from the resource allocation. If the Quality of Experience  (QoE) is improved, they continue to follow the hint. Otherwise they revert back to their original allocation or randomly select a new channel.   Overall the users gain from participating. If the system is sabotages, they will temporarily select a channel that reduces their QoE. However, they will rapidly revert back to an expected  QoE they would have without participating in the system.  If the system is not sabotaged, the QoE is improved.
 As part of a cyber-attack, external force can inject massive location messages in order to try to disrupt the system. In the simulations, we have shown that the system can handle large volumes of traffic. In order to be successful, order of magnitude larger volumes will potentially be needed in order to disrupt the system.

\section{Other proximity service applications}
\label{other}
The main purpose of this paper is to introduce how a DSA system can be implemented for  IoT and M2M devices as discussed in the previous sections. There may be other applications too presented in the following, but these have not been evaluated thoroughly.

\subsection{Context-Aware Services for roaming nodes }
\label{CAS}

To adapt the services to the context of the user considering his environment, is very popular and involves using the knowledge about the location of the user to offer services.  As an example, once the location is known, geospatial databases are often used to answer queries from mobile computer users for the nearest post office, best realtor, directions to the airport, and so forth. Considering where the mobile user has been and projecting in which direction he is going, could offer yet, more services. For this purpose the Discovery Agent presented in this paper will be most useful. A common way to implement publish/subscribe M2M services is by using Web services. The OASIS standard WS-Notification can be leveraged for this purpose, as it describes how topics can be used for subscription using XML schemas.  How to extend this scheme to locations as a topic is not straight forward especially for mobile nodes for which the topics must be organized in areas and new topics need to be addressed following the route. An easier way is to restrict the search for topics among the candidates on the candidate node list provided by the Discovery Agent.

\subsection{Contextual Services for IoT beween different organizations}
\label{Context}
As long as the devices are registred and managed by one organization, there exist lists of devices and the need for coordination is less. Between devices belonging to different groups like that which is found e.g. in a smart city, there is already many devices which may benefit from cooperation to make better judgments which make the life easier for the occupants. Motion detectors used for burglar alarms can  be shared for switching on the light in rooms and shut it off as there is no movement. Smoke detectors for fire alarm may be shared with systems for illumination by switching on the light when an alarm is triggered. Water leakage triggered by a dishwasher may also result in that the light is switched on etc.
All these sensors and more, can be monitored  and managed remotely by mobile devices. To make the information of the sensors available to all other in a local network, the can be connected by wire or by radio to Internet. To make them aware of each other and provide a IP connectivity matrix a Discovery protocol is must useful.

\subsection{Handover }
\label{Handover}

Once the nodes move withing the coverage of fixed access points, there is a need of hand-over to ensure uninterrupted services. For mobiles,3GPP compliant mobile networks (3GPP in Release 8) \cite{sachs2010access} uses an system architecture called Access Network Discovery and Selection Function (ANDSF) used by the evolved packet core (EPC) network. This enables not only user equipment like 3GPP access networks (such as HSPA or LTE) , but also  non-3GPP access networks – such as Wi-Fi or WIMAX to be discovered by radio and connect to each other. A similar functionality for Wi-Fi is IEEE 802.11u. 802.11u was developed to effectively automate how devices connect to available Wi-Fi networks. 802.11u enables Wi-Fi hotspots to advertise their capabilities and then allows devices to connect to them automatically rather than requiring the end user to manually select an SSID. One of the challenges when including Wi-Fi traffic into the 3GPP network is that Wi-Fi traffic must  be connected to the EPC network via secure IP tunnels (GRE, IPSec. This makes this interconnection challenging to implement.

In IEEE 802.21 Information Service is specified to allow network entities to discover information 
that influences the selection of appropriate networks during handovers. Since this structure is based upon a centralized concept, this information may supplement the  Candidate Node List for a given geographical area and shared directly with surrounding nodes to reduce traffic to the Information Service server.


The ANR \cite{sesia2009lte} function relies on cells broadcasting their identity. 
In order to minimize the time of no connection during handover, it is very useful to be able to configure the operating frequency prior to the new link establishment. 

In the FP7 EU project MOTO, Wi-Fi technology is considered for offloading mobile networks \cite{6686527} .  A similar effort is to use Heterogeneous Network (Hetnet)   based on vendor 3GPP-standardized and coordinated radio network with integrated Wi-Fi and  traffic management. In spite of this, there is a strong unwillingness among operators to introduce features which may result in loss of paid services. To enable handover from a LTE network to a Wi-Fi system may therefore be difficult unless the operators are the owners of the Wi-Fi system as a part of the 3GPP architecture. This reluctance exists even if a handover to a Wi-Fi net is in the interest of the operator to relieve the operator network which is fully loaded.

\subsection{Many small base stations} 
When femtocells or Hotspots are deployed in large quantities , it will be practically impossible to do careful radio-planning. For this purpose, the concept of Self Organizing Networks (SON) has been introduced in  3GPP Release 8,  to make planning, configuration, management, optimization and healing of mobile radio access networks simpler and faster. This initiative can enjoy the automatic Distributed Discovery System proposed. Since these are normally owned and operated by an operator, the members are authenticated and the members can trusted.

To enable the mobile network to offer more capacity to many users even if the spectrum is limited, it is necessary to make the number of users per base station small. Ultimately there will only be one user per base station and this user does not need to share the bandwidth with other users. This base station is called a femtocell and a network of very small and large base station is called HETerogenous Networks (HET-NET)  \cite{6354282} in LTE  and Hotspot in Wi-Fi. Hotspot 2.0 is an attempt to automate network discovery, registration, provisioning, and network connectivity, which are manual steps today when a user connects to a given Wi-Fi hotspot.

A problem with the introduction of pico-cells in LTE comes from cell
association protocols. Traditionally the user connects to the cell with the highest
signal power to achieve maximum user performance, but in the case of Hetnet s
this can degrade the overall throughput because the pico-cells have much lower
transmit power than the macro cells.This means that the macro cell serves more
UEs than it should and the pico-cell becomes under utilized because of small
coverage area. This problem can be solved by creating Neighbor Relation Tables produced from information obtained through the back haul network rather then radio, using the results of  the Discovery system architecture proposed in this paper.

\section{Discussion and Conclusion} 

A Distributed Discovery System has been described which is believed to be well suited for IoT and M2M devices and particular those owned and operated by individuals. 
It is our goal with this effort to initiate further modules like candidate Resource Allocation components to be implemented at tested a shared and finally standardized.  Another example is to spur new ideas related the shared frequency bands and in particular the White Space.

The advantage of this gradual implementation approach is that no operator need to be chosen and no common infrastructure is needed- it may be implemented by two neighbors and grow organically as the usefulness of the system is accepted.

\bibliographystyle{spmpsci}
\bibliography{BibTeX1}

\end{document}